\documentstyle{mn}
\input psfig

\newcommand{\PPF}{partition probability function}
\newcount\eqnno \eqnno=0 
\def\numeqn{\global\advance\eqnno by 1 \eqno(\the\eqnno)} 
\def\etal{et al. }

\title[Forest of merger history trees]
{The forest of merger history trees associated with the  
formation of dark matter halos}

\author[R. K. Sheth \& G. Lemson]
{Ravi K. Sheth and Gerard Lemson\\
MPI-Astrophysik, K. Schwarzschildstr. 1, 85748 Garching, Germany\\
\smallskip
Email: sheth@mpa-garching.mpg.de, lemson@mpa-garching.mpg.de\\
}

\date{Submitted May 1998}

\begin{document}

\maketitle

\begin{abstract}
We describe a simple efficient algorithm that allows one to construct 
Monte-Carlo realizations of merger histories of dark matter halos. 
The algorithm is motivated by the excursion set model 
(Bond et al. 1991) for the conditional and unconditional halo mass 
functions.  The forest of trees constructed using this algorithm 
depends on the underlying power spectrum.  For Poisson or 
white-noise initial power-spectra, the forest has exactly the 
same properties as the ensemble of trees described by Sheth (1996) 
and Sheth \& Pitman (1997).  In this case, many ensemble averaged 
higher order statistics of the tree distribution can be computed 
analytically.  For Gaussian initial conditions with more general 
power-spectra, mean properties of the ensemble closely resemble 
the mean properties expected from the excursion set approach.  
Various statistical quantities associated with the trees constructed 
using our algorithm are in good agreement with what is measured 
in numerical simulations of hierarchical gravitational clustering.  

\end{abstract} 
\maketitle

\section{Introduction}
It is widely believed that the massive dark matter halos which exist 
today have grown from small initially Gaussian density fluctuations.  
Many questions of astrophysical interest can be addressed using 
Monte--Carlo realizations of the merger histories of such massive 
dark matter halos.  An efficient method for generating Monte-Carlo 
realizations of the clustering process is essential for addressing 
such questions (e.g. Kauffmann \& White 1993).  
Ideally, such an algorithm should produce results that are consistent 
with other known properties of the clustering process.  

In this paper, we develop an algorithm which is motivated by known 
excursion set results (Bond et al. 1991; Lacey \& Cole 1993).  
For example, these earlier results provide expressions for the mean 
number of halos of mass $m$ that are later incorporated into a 
larger halo of mass $M>m$.  For the special case of clustering from 
Poisson, or white-noise initial conditions, a model for the higher 
order moments of this subclump distribution exists (Sheth 1996; 
Sheth \& Pitman 1997).  Appendix~\ref{ppf} of this paper shows that 
these higher order moments are consistent with the fact that mutually 
disconnected regions in Poisson or white-noise Gaussian distributions 
are mutually independent.  
Section~\ref{pics} describes an algorithm which uses 
this fact to generate a forest of merger trees.  The algorithm 
reproduces the mean and higher-order Poisson and white-noise results 
exactly.  Although the algorithm is not of binary split type, 
trees constructed using binary splits provide good approximations 
to ours (Section~\ref{bsplit}).  

Unfortunately, previous work does not provide expressions for the 
higher order moments of the subclump distribution for other initial 
power spectra (e.g. Sheth \& Pitman 1997).  
Section~\ref{mctree} shows that, to describe the trees associated 
with arbitrary Gaussian initial conditions, a simple modification 
of our white-noise algorithm produces good approximations to the 
known (excursion set) mean values.  
Section~\ref{scfsim} shows that, for a range of power spectra of current 
interest, our algorithm provides higher order distributions that are 
similar to those measured in numerical simulations of hierarchical 
gravitational clustering.  A final section summarizes our results.  
It argues that, in addition to allowing one to generate a forest 
of merger history trees, our results are also useful for studying 
the spatial distribution of dark matter halos.  

\section{Poisson initial conditions}\label{pics}
The first part of this section summarizes the relevant results of 
Sheth (1996), where details of all the arguments are given.  
The second part shows how these earlier results can be used to 
derive an algorithm for partitioning halos into subclumps.  The 
algebra for some of the arguments in this section is given in 
Appendices~\ref{ppf} and~\ref{permute}.  

\subsection{Excursion set and branching process results}
The evolution of clustering from an initially Poisson distribution 
has been studied using the excursion set (Epstein 1983) and 
cloud-in-cloud formalisms (Sheth 1995).  In these analyses, 
all matter is bound into clumps, sometimes called halos.  
The probability that a clump has $m$ associated particles is 
\begin{equation}
\eta(m,b) = {(mb)^{m-1} {\rm e}^{-mb}\over m!} ,
\label{borel}
\end{equation}
where $m\ge 1$ and $0\le b\le 1$.  
Equation~(\ref{borel}) is known as the Borel distribution 
(Borel 1942).  The probability $f(m,b)$ that a randomly 
chosen particle belongs to an $m$-clump is given by multiplying 
$\eta(m,b)$ by $m$, and dividing by the mean clump size 
$\sum_m m\,\eta(m,b) = 1/(1-b)$:
\begin{equation}
f(m,b) = m\,(1-b)\,\eta(m,b).
\label{fkb}
\end{equation}
Since all matter is in clumps, and $f(m,b)$ is the fraction of mass 
in $m$ clumps, 
\begin{equation}
\sum_{m=1}^\infty f(m,b) = 1.
\end{equation}
The evolution of clustering is described by the evolution of the $b$ 
parameter, which is the ratio of the average density in the universe 
to the critical overdensity that is required for an object to collapse 
and virialize at cosmological time $t$.  That is, 
\begin{equation}
b(t) = {1\over 1+\delta_{\rm c}(t)}\qquad{\rm where}
\quad \delta_{\rm c}(t)\propto {1.686\over a(t)},
\label{bdt}
\end{equation}
$t$ denotes cosmological time, and $a(t)$ is the 
factor by which the universe has expanded since some fiducial initial epoch, 
say, $t_i$.  Notice that the critical density decreases as the universe 
expands, so that $b$ increases with expansion.  Since $b$ 
increases monotonically as cosmological time increases, it can 
also be treated as a pseudo-time variable.  
Thus, $b$ plays two roles: it can be thought of as describing a 
spatial density, or it can be thought of as describing time.  
Therefore, $f(m,b)$ can be thought of as the fraction of the total mass
that is in $m$-clumps at time $b$, or it can be thought of as the 
fraction of the total mass that is associated with regions within 
which the average overdensity is given by $b$.  

The excursion set description also allows one to compute the probability 
that a randomly chosen particle of a clump having exactly $M$ particles  
at the time corresponding to $b_0$ was a member of an $m$-clump at the 
earlier epoch $b_1<b_0$.  This probability is 
\begin{eqnarray}
f(m,b_1|M,b_0) &=& M\left(1-{b_1\over b_0}\right)\,
{M\choose m}\,{m^m\over M^M}\nonumber \\
&&\ \times\ \left({b_1\over b_0}\right)^{m-1}
\left[M-m{b_1\over b_0}\right]^{M-m-1}\!\!,
\label{fjk}
\end{eqnarray}
where $1\le m\le M$ and $0\le b_1/b_0\le 1$  (Sheth 1995).  
This implies that the average number of progenitor $m$-subclumps 
identified at $b_1$ that are later (at $b_0>b_1$) in an $M$-clump is 
\begin{equation}
N(m,b_1|M,b_0) = (M/m)\,f(m,b_1|M,b_0).   
\label{njk}
\end{equation}
Although the excursion set formalism provides information about the 
mean number of progenitor subclumps of any given parent clump, 
it does not provide a complete description of the 
entire merger history tree (cf. Fig.~1 in Sheth \& Pitman 1997).  

Let $p({\bmath n},b_1|M,b_0)$, with $\bmath{n} = (n_1,\cdots ,n_M)$, 
$\sum_{j=1}^M n_j = n$ and $\sum_{j=1}^M j\,n_j = M$, denote the 
probability that an $M$-clump at the epoch $b_0$ was previously in 
$n$ subclumps at the epoch $b_1$, of which there were $n_1$ single 
particle subclumps, $n_2$ subclumps with exactly two 
particles, $n_j$ subclumps with exactly $j$ particles, and so on.  
Then the branching process model of Sheth (1996) yields the formula
\begin{equation}
p({\bmath n},b_1|M,b_0) =
{(Mb_{01})^{n-1}{\rm e}^{-Mb_{01}}\over \eta(M,b_0)}\,
\prod_{j=1}^M {\eta(j,b_1)^{n_j}\over n_j!},
\label{prtfnc}
\end{equation}
where $b_{01}=(b_0-b_1)$.  Sheth \& Pitman (1997) showed that 
this same formula follows from another construction, which 
they termed the additive coalescent.  
Appendix~\ref{ppf} of this paper shows that this formula is 
consistent with the assumption that mutually disconnected regions 
within a Poisson distribution are mutually independent.  

Notice that equation~(\ref{prtfnc}) depends only on the ratio 
$b_1/b_0$, and not on the actual values of $b_1$ or $b_0$.  
This dependence on the ratio $b_1/b_0$ implies that the merger 
histories of $M$-clumps formed at the epochs $b_0$ and $b_0'$ 
are related to each other simply by rescaling the time variable.  
By taking relevant sums over the partition probability function, one 
can show explicitly that the branching process expressions for the  
clump size distribution and the merger probabilities $f(m|M)$ are 
identical to the excursion set results (equations~\ref{borel} 
and~\ref{fjk}).  
However, the \PPF\ also allows one to compute quantities that 
cannot be estimated using the excursion set formalism.  
For example, the excursion set approach says that the mean number 
of $m$-subclumps per $M$-clump is $(M/m)\,f(m|M)$. 
The \PPF\ allows one to compute the higher order moments of the 
$n_j$ distribution also.  For example, 
\begin{eqnarray}
&&\!\!\!\!\!\!\!\!\!\!\!\!
\left\langle\!{n_i!\over (n_i-\alpha)!}{n_j!\over
(n_j-\beta)!},b_1\,\Biggl\vert\,M,b_0\!\right\rangle \nonumber \\
&&\ =\ \Bigl[Mb_{01}\Bigr]^{\alpha + \beta} \ 
{\eta^\alpha(i,b_1)\,\eta^\beta(j,b_1)\ \eta(R,B)\over\eta(M,b_0)} ,
\ \ {\rm if}\ R>0\nonumber \\
&&\ =\ \Bigl[Mb_{01}\Bigr]^{\alpha + \beta - 1} \ 
{\eta^\alpha(i,b_1)\,\eta^\beta(j,b_1)\over\eta(M,b_0)} ,
\ \ {\rm if}\ R=0,
\label{facij}
\end{eqnarray} 
where $b_{01} \equiv (b_0-b_1)$ and 
\begin{equation}
RB\equiv (M-\alpha i- \beta j)B = Mb_0-\alpha ib_1 - \beta jb_1 .
\end{equation}
When $\alpha=1$ and $\beta=0$ this is equivalent to the relation 
obtained from equation~(\ref{fjk}).  When $\alpha=\beta=1$, 
equation~(\ref{facij}) gives an estimate of the cross-correlation 
between $i$- and $j$-subclumps that are within the same $M$-clump.  

\subsection{The partition algorithm}\label{partalgo}
The partition algorithm described below follows from the fact 
(cf. Appendix~\ref{ppf}) that equation~(\ref{prtfnc}) is consistent 
with the assumption that mutually disconnected regions within a 
Poisson distribution are mutually independent.  

Imagine partitioning an $M$-clump up into subclumps by choosing first 
one subclump, then choosing a second from the mass that remains, 
and so on until all the mass $M$ has been assigned to subclumps.  
Start by choosing a random one of the $M$ particles.  
Equation~(\ref{fjk}) says that, on average, the chosen particle will 
be a member of an $m_1$-subclump with probability $f(m_1,b_1|M,b_0)$.  
Since $b$ is a density variable, the halo $M$ can be thought of 
as being a region of size $V_M = Mb_0/\bar n$, where 
$\bar n$ denotes the average density.  If the first subclump 
has mass $m_1$, then it is associated with a volume 
$V_{m_1}=m_1b_1/\bar n$ within $V_M$.  The remaining 
$R_1\equiv (M-m_1)$ particles must be distributed randomly (because 
the initial distribution is Poisson) within the remaining volume 
$V_M-V_{m_1}$.  Let $\delta^{(1)}$ denote the average density in 
this remaining volume, and recall (equation~\ref{bdt}) that it can 
be represented by some $b^{(1)}\equiv 1/(1+\delta^{(1)})$.  Then   
\begin{equation}
{\bar n\over b^{(1)}}\equiv {M-m_1\over V_M - V_{m_1}} \ \ \ \ 
{\rm and\ so}\ \ 
R_1\,(b^{(1)} - b_1) = Mb_{01}.
\label{bprime}
\end{equation}
Now the second subclump of the $(M,b_0)$-halo must be chosen from 
among the remaining $R_1=M-m_1$ particles.  
Suppose that the probability that this second subclump has exactly 
$m_2$ particles is given by the same rule as before, namely, 
by $f(m_2,b_1|R_1,b^{(1)})$.  If the second subclump has mass $m_2$, 
then there are $R_2\equiv (M-m_1-m_2)$ remaining particles, 
they are randomly distributed within $(V_M-V_{m_1}-V_{m_2})$, 
where $V_{m_2} = m_2b_1/\bar n$, and the average density within 
this remaining region is parametrized by some $b^{(2)}$, defined 
analogously to equation~(\ref{bprime}).  
Suppose that the third subclump has exactly $m_3$ particles with 
probability $f(m_3,b_1|R_2,b^{(2)})$, and so on.  

Let $m_i$ denote the number of particles in the $i$th subclump.  
Let $R_i$ denote the mass remaining after the $i$th pick:  
\begin{equation} 
R_i \equiv M-\sum_{j=1}^i m_j,\ \ \ \ {\rm and}\ \ \ R_0 \equiv M, 
\end{equation}
and parametrize the average density in the volume that remains 
after the $i$th pick by $b^{(i)}$.  Then   
\begin{displaymath}
{1\over b^{(i)}} = {R_i\over \bar nV_M-\sum_{j=1}^i \bar nV_{m_j}}
= {R_i\over Mb_{01} + R_ib_1}.
\end{displaymath}
Notice that $1/b^{(0)} = R_0/(Mb_{01} + R_0b_1) = 1/b_0$ as required, 
and that 
\begin{equation}
R_i\,(b^{(i)} - b_1) = 
R_i\,\left({Mb_{01} + R_ib_1\over R_i} - b_1\right) 
= Mb_{01}
\end{equation}
is independent of $i$.  In this notation, equation~(\ref{fjk}) 
becomes 
\begin{eqnarray}
&&\!\!\!\!\!\!\!\!\!\!\!\!\!\!\! 
f(m_i,b_1|R_{i-1},b^{(i-1)}) \nonumber \\
&=& m_i\,(b^{(i-1)}-b_1)\, 
{\eta(m_i,b_1)\,\eta(R_i,b^{(i)})\over \eta(R_{i-1},b^{(i-1)})},
\nonumber \\
&=& {m_i\over R_{i-1}}\,Mb_{01}\,
{\eta(m_i,b_1)\,\eta(R_i,b^{(i)})\over \eta(R_{i-1},b^{(i-1)})}
\ \ \ \ {\rm if}\ \ 1\le m_i<R_{i-1} \nonumber \\
&&\!\!\!\!\!\!\!\!\!\!\!\!\!\!\! 
{\rm and\ it\ is} \nonumber \\ 
&=& {\eta(m_i,b_1)\over \eta(R_{i-1},b^{(i-1)})}\ 
{\rm e}^{-R_{i-1}(b^{(i-1)} - b_1)} \nonumber \\
&=& {\eta(m_i,b_1)\over \eta(R_{i-1},b^{(i-1)})}\ 
{\rm e}^{-Mb_{01}}\ \ \ \ \ \ \ {\rm if}\ \ m_i=R_{i-1}.
\label{factrd}
\end{eqnarray}
The probability that after $n$ picks all $M$ particles have 
been assigned to a subclump (i.e., $m_n=R_{n-1}$ and $R_n=0$), is 
simply the product of the individual $n$ choices above.  
If ${\bmath m}$ denotes the set $(m_1,\cdots,m_n)$, then 
\begin{eqnarray}
p({\bmath m}|M) &=& 
\prod_{i=1}^n f(m_i,b_1|R_{i-1},b^{(i-1)}) \nonumber \\
&=& {(Mb_{01})^{n-1}\,{\rm e}^{-Mb_{01}}\over \eta(M,b_0)}
\prod_{i=1}^n {m_i\over R_{i-1}}\,\eta(m_i,b_1) ,
\label{prod}
\end{eqnarray}
where the final expression uses the fact that $R_0=M$, 
$b^{(0)} = b_0$, and $m_n=R_{n-1}$.  

Equation~(\ref{prod}) describes the probability that $M$ was 
partitioned into these $n$ subclumps in the order $(m_1,\cdots,m_n)$.  
The probability that the sequence of picks resulted in the 
set $(m_1,\cdots,m_n)$, whatever the order of the $m_i$, is given 
by summing equation~(\ref{prod}) over all $n!$ permutations of 
the $n$ picks $m_i$.  Let $\pi[{\bmath m}]$ denote this set of 
permutations.  Then 
\begin{eqnarray}
\sum_{\pi[{\bmath m}]} p({\bmath m}|M) &=& 
\sum_{\pi[{\bmath m}]} 
{(Mb_{01})^{n-1}\,{\rm e}^{-Mb_{01}}\over \eta(M,b_0)} \nonumber \\
&&\ \ \times\ 
\left(\prod_{i=1}^n {m_i\over R_{i-1}}\right)
\left(\prod_{i=1}^n\eta(m_i,b_1)\right),
\end{eqnarray}
since, for a given ${\bmath m}$, all terms in each permutation are 
the same, except for the product $\prod_{i=1}^n (m_i/R_{i-1})$.  
So, the sum over permutations becomes 
\begin{eqnarray}
\sum_{\pi[{\bmath m}]} p({\bmath m}|M) 
&=& {(Mb_{01})^{n-1}\,{\rm e}^{-Mb_{01}}\over \eta(M,b_0)}
\left(\prod_{i=1}^n\eta(m_i,b_1)\right) \nonumber \\
&& \qquad\qquad\times\ \ \sum_{\pi[{\bmath m}]} 
\left(\prod_{i=1}^n {m_i\over R_{i-1}}\right) .
\label{perms}
\end{eqnarray}
A little thought shows that the contribution from the terms in 
the sum is unity.  Appendix~\ref{permute} shows this explicitly.  
Thus, 
\begin{equation}
\sum_{\pi[{\bmath m}]} p({\bmath m}|M) = 
{(Mb_{01})^{n-1}\,{\rm e}^{-Mb_{01}}\over \eta(M,b_0)}
\prod_{i=1}^n \eta(m_i,b_1).
\label{simpler}
\end{equation}
Now, if all particles are indistinguishable, then subclumps that 
have the same number of particles are indistinguishable from each 
other, so these permutations should not be counted separately.  
We can account for this by dividing the product above 
by $\prod_{i=1}^M n_i!$, where $n_i$ denotes the number of 
$i$-subclumps in the partition, $\sum_i n_i = n$, and 
$\sum_i i\,n_i=M$.  Thus, the probability that $M$ was partitioned 
into ${\bmath n}$ is the same as that given by 
equation~(\ref{prtfnc}).  

This decomposition suggests the following algorithm for 
partitioning an $(M,b_0)$-halo into $b_1$-subhalos.  
\begin{verbatim}
  Initialize:  specify b1, b0, M, and set R = M.  
    Set b = M(b0-b1)/R + b1.  
    Choose m with probability f(m,b1|R,b).  
    Set R = R-m.
  Iterate until R = 0. 
\end{verbatim}
The list of $m$'s so obtained gives one possible partition 
of $M$ into $b_1$-subclumps.  
Repeat to generate an ensemble of such partitions.
The previous paragraphs show that this algorithm will produce 
partitions with the same ensemble properties as the branching
process (equation~\ref{prtfnc}).  

Notice that, other than requiring that $0\le b_1/b_0\le 1$, 
there is no other constraint on $b_1/b_0$.  This means that 
the algorithm constructs partitions with the correct ensemble 
properties in one time step.  However, one is often interested 
in following the history of a halo as it becomes partitioned into 
smaller pieces through many small time steps.  
It is simple to use this partition algorithm to construct such a 
merger history tree.  Given $M$ and $b_0$, choose $b_1<b_0$.  
Use the algorithm above to partition $M$ into $b_1$-subclumps.  
Call this partition ${\bmath m}$.  
Choose $b_2<b_1$.  Partition each subclump $m$ of ${\bmath m}$ 
into its constituent subclumps using the algorithm above, 
with the initial parameters $b_1\to b_2$, $b_0\to b_1$, $M\to m$ 
and set $R=m$.  Then choose $b_3<b_2$, and so on.  The results of 
Sheth \& Pitman (1997) and Sheth (1998) show that this many-step 
algorithm is guaranteed to give the same ensemble of partitions 
as the one-step algorithm described in detail above.  

\section{Gaussian initial conditions}\label{gics}
This section shows that the partition algorithm of the previous 
section can easily be generalized to partition halos associated 
with Gaussian initial fluctuations.  

Let $\delta(t)$ denote the critical density required for a 
spherical perturbation to collapse by time $t$ 
(cf. equation~\ref{bdt}).  Thus, $\delta$ is both an overdensity 
variable and a pseudo-time variable, as was $b$ in the 
previous section.  Also, let $\sigma^2(M)$ denote the variance 
in the initial density fluctuation field when smoothed over 
regions that contain mass $M$ on average.  The variance depends 
on $M$ in a way that depends on the power-spectrum; we will only 
be interested in initial power-spectra for which $S(M)$ decreases 
monotonically as $M$ increases (e.g. Press \& Schechter 1974).  

In the excursion set approach (Bond \etal 1991; Lacey \& Cole 1993), 
the fraction of mass that is in halos which have mass in the range 
${\rm d}m$ about $m$ at cosmological time $t$ is 
\begin{equation}
f(m,\delta)\,{\rm d}m = \sqrt{\delta^2\over 2\pi S}\,
\exp\left(-{\delta^2\over 2S}\right)\,
{{\rm d}S\over S},
\label{fmps}
\end{equation}
where $S=\sigma^2(m)$ and $\delta$ decreases as cosmological time 
increases, as described above.  
The average number density of such halos is 
\begin{equation}
n(m,\delta)\,{\rm d}m = (\bar\rho/m)\,f(m,\delta)\,{\rm d}m,
\end{equation}
where $\bar\rho$ denotes the average density.  

Similarly, consider a halo that has mass $M$ at time $\delta_0$.  
The fraction of the mass of this halo that was in subhalos of 
mass $m$ at the earlier time $\delta_1>\delta_0$ is 
\begin{equation}
f(m,\delta_1|M,\delta_0)\,{\rm d}m = 
{\sqrt{\nu}\,{\rm e}^{-\nu}\over\sqrt{\pi}}\,
{{\rm d}\nu\over\nu},\ \ \ {\rm provided}\ \ m\le M,
\label{fmMps}
\end{equation}
\begin{displaymath}
{\rm where}\ \nu = {(\delta_1-\delta_0)^2\over 2 (s-S)},\ 
s=\sigma^2(m),\ {\rm and}\ S=\sigma^2(M).
\end{displaymath}  
Thus, the mean number of $m$ halos within an $M$ halo is  
\begin{equation}
N(m,\delta_1|M,\delta_0)\,{\rm d}m = 
\left({M\over m}\right)\,f(m,\delta_1|M,\delta_0)\,{\rm d}m
\label{nmM}
\end{equation}
(e.g. Lacey \& Cole 1993).

\subsection{White-noise initial conditions}
For white-noise, $\sigma^2(M)\propto 1/M$.  
In the limit of small $\delta$, which corresponds to the 
$b\to 1$ limit (cf. equation~\ref{bdt}), 
Stirling's approximation for the factorials reduces all the Poisson 
excursion set statements (equations~\ref{borel} and~\ref{fjk}) to the 
corresponding excursion set statements (\ref{fmps} and \ref{fmMps}) 
for $n=0$ white-noise initial conditions (Sheth 1995, 1996).  
Notice that the $b_1/b_0$ dependence of the Poisson expressions 
translates to a dependence on $(\delta_1-\delta_0)$, rather than 
the actual values of $\delta_1$ or $\delta_0$ themselves.  
This means that $f(m,\delta_1|M,\delta_0)$ can be written as 
$f(m|M,D)$, where $D = (\delta_1-\delta_0)$.  Furthermore, 
to lowest order in $\delta$, $f(m|M)$ can be factored 
similarly to (\ref{factrd}).  That is, just as in the Poisson 
case, $M$ occupies a region $V$, such that 
$M/V = \bar\rho(1+\delta_0)$, where $\bar\rho$ is the background 
density.  Similarly, the subhalo $m$ occupies $v$ within $V$, 
with $m/v = \bar\rho (1+\delta_1)$.  
Thus, the density in the remaining volume is 
\begin{equation}
{M-m\over V-v} = \bar\rho(1+\delta'), \ \ \ {\rm so}\ \ 
(\delta_1-\delta') = {(\delta_1-\delta_0)\over 1-(m/M)}
\label{dprime}
\end{equation}
to lowest order in the $\delta$ terms.  Finally, since 
disconnected volumes are mutually independent in the 
white-noise case, just as in the Poisson case, the same 
partition algorithm that worked there will also work here:  
\begin{verbatim}
  Initialize:  specify d1, d0, M, and set R = M.  
    Set D = (d1-d0)/(R/M). 
    Choose m with probability f(m|R,D).  
    Set R = R-m.  
  Iterate until R is as small as desired.  
\end{verbatim}
A simple transformation of equation~(\ref{fmMps}) shows that 
to choose $m$ according to $f(m|R,D)$ one needs to generate a 
Gaussian random variable.  
There are efficient ways to do this, so this algorithm is fast.  

This algorithm allows one to generate partitions of $M$ whose 
distributional properties are described by the white-noise analogue 
of equation~(\ref{prtfnc}).  As before, there is no restriction 
on the size of the `time step':  $(\delta_1-\delta_0)>0$ can be set 
as large as desired (although the algorithm is slower for larger 
time steps, since, for large $\delta_1-\delta_0$  most partitions 
are into many tiny pieces).  And again, constructing a merger 
history tree by embedding this partition algorithm in a loop 
over time-steps is straightforward.  In either case, the algorithm 
generates subclump distributions whose mean values are exactly 
the same as those required by the excursion set approach 
(equation~\ref{fmMps}).  
The higher order moments of the subclump distribution are 
given by applying Stirling's approximation to the 
corresponding Poisson expressions, and taking the 
$\delta\ll 1$ limit.  For example, the cross-correlation 
between $m_1$ and $m_2$ halos that are within the same $M$ 
halo is 
\begin{equation}
c(12|0) = N(m_1,\delta_1|M,\delta_0)\,N(m_2,\delta_1|M-m_1,\delta'),
\label{c120}
\end{equation}
where $N(m|M)$ was defined by equation~(\ref{nmM}) and $\delta'$ was 
defined by equation~(\ref{dprime}).  
For white-noise $c(12|0) = c(21|0)$.

\subsection{Arbitrary Gaussian initial conditions}
The assumption that disconnected volumes are mutually independent
is wrong when the initial conditions differ from white-noise.  
Therefore, it is not obvious that the algorithm above should be 
used to generate partitions for arbitrary initial conditions.  
On the other hand, for arbitrary initial conditions, nothing 
is known about the higher order moments of the subclump 
distribution; only the mean is known from the excursion set 
approach.  Below, we take two different approaches to constructing 
a partition algorithm for initial conditions that are not 
white-noise.  Which algorithm to use is determined by comparison 
with numerical simulations in section~\ref{scfsim}.  

The first approach is motivated by the observation that, 
when expressed as functions of the variance rather than the 
mass, all excursion set quantities are independent of power 
spectrum.  This suggests that, with a suitable transformation, 
we should be able to use the algorithm above, since it is known 
to be exact for the white-noise spectrum.  We do this as follows.  
Run the algorithm assuming a white-noise power-spectrum.  
However, treat each chosen $m$ not as a subclump having mass $m$, 
but as a region of volume $v/V= m/M$ containing mass $m$, that 
is populated by some number $\nu$ of objects that all have the same 
mass $\mu$.  Thus, $\nu=m/\mu$.  The value of $\mu$ is got by 
requiring that $\sigma^2(\mu)$ equals the white-noise value, 
namely $\sigma^2(\mu)= 1/m$.  To illustrate, suppose that 
$\sigma^2(\mu)=\mu^{-\alpha}$.  Then $\nu$ is 
$m\, \sigma^2(\mu)^{1/\alpha} = m^{(\alpha-1)/\alpha}$.  
For white noise, $\alpha=1$, so $\nu=1$; thus, the region $v$ 
contains exactly one halo as in the previous subsection.  
For general $\alpha$, $\nu$ is neither unity nor even integer.  
Nevertheless, this approach generates partitions which are 
guaranteed to have the excursion set value in the mean, but 
which, as a result of the fact that $\mu$ halos always come in 
groups of $\nu$, may have unnaturally high values for the higher 
order moments.  Since this approach simply uses the white-noise 
algorithm, the higher order moments can be calculated 
analytically.  For example, if $\sigma^2(\mu)= \mu^{-\alpha}$, 
then the cross-correlation between halos is 
\begin{equation}
C(12|0) = \left({m_1\over M}{m_2\over M}\right)^{{(\alpha-1)\over\alpha}}
c(12|0)
\label{c120p}
\end{equation}
where $c(12|0)$ is given by equation~(\ref{c120}) but with the 
white-noise power spectrum.

The second approach is to assume, with no real justification, that 
the algorithm of the previous subsection can be used directly.  
In this case, the ensemble of partitions will differ for different 
power-spectra only because $\sigma^2(M)$ in~(\ref{fmMps}) depends 
on power-spectrum.  Since, in general, disjoint volumes are not 
independent, this means, of course, that there is no longer any 
guarantee that the algorithm will generate the excursion set mean 
values, either when used as a one step partition algorithm, or when 
the partition algorithm has been embedded in a loop over timesteps.  
Another way to see this is to note that $f(m|M)$ factors 
conveniently only for a white-noise power-spectrum.  
Since there is no guarantee that the algorithm will produce the 
excursion set statements in the mean, we also have no real analytic 
estimate for the higher order moments associated with this algorithm. 
Later, when we show the results using either the ensemble of one-step 
partitions or multi-step merger trees associated with this approach, 
we will compare the Monte-Carlo results with the formulae given in 
the previous subsection (\ref{fmMps}, \ref{nmM} and~\ref{c120}), 
but with the relation $\sigma^2(M)$ depending on power spectrum.  

\subsection{Relation to binary split models}\label{bsplit}
Binary split algorithms based on the excursion set approach 
have been discussed by Lacey \& Cole (1993) and by 
Sheth \& Pitman (1997).  
In the limit $\delta_1-\delta_0\equiv \delta_{10}\ll 1$, the 
one-step partition algorithm discussed above will produce partitions 
in which most of the mass is in the largest two pieces.  
To see this, recall that each choice in the partition process is 
of some $s-S(R) = (\delta_{10}/x)^2$, where $x$ is a 
Gaussian random variable with zero mean and unit variance.  
Since $x$ is Gaussian, most of time, $s-S(R)\approx \delta_{10}^2$.  
If $\delta_{10}\ll 1$, then $s\approx S(R)$ after most draws of 
the random number $x$.  This means that, most of the time, the 
first choice is of a subhalo with most of the mass of $M$.  
Iterating this argument, the second choice is of a subhalo which 
contains most of the remaining mass.  
Since a given range in $s-S$ correponds to a smaller mass 
range for white-noise than for power-spectra with more negative 
slopes, the above is more true for white-noise than for 
these other power-spectra.  Therefore, at least for white-noise, 
a binary split algorithm, with $f(m|M)$ for the split rule, and 
with succesive splits spaced at $\Delta\delta=\delta_{10}\ll 1$, 
should provide a good approximation to the merger trees produced 
here.  A different approach to just such a binary split algorithm 
is described by Sheth \& Pitman (1997).  They also discuss why 
binary split algorithms that are consistent with the excursion 
set approach are easier to construct in the white-noise 
case than in the general case.  

\begin{figure*}
\centering
\mbox{\psfig{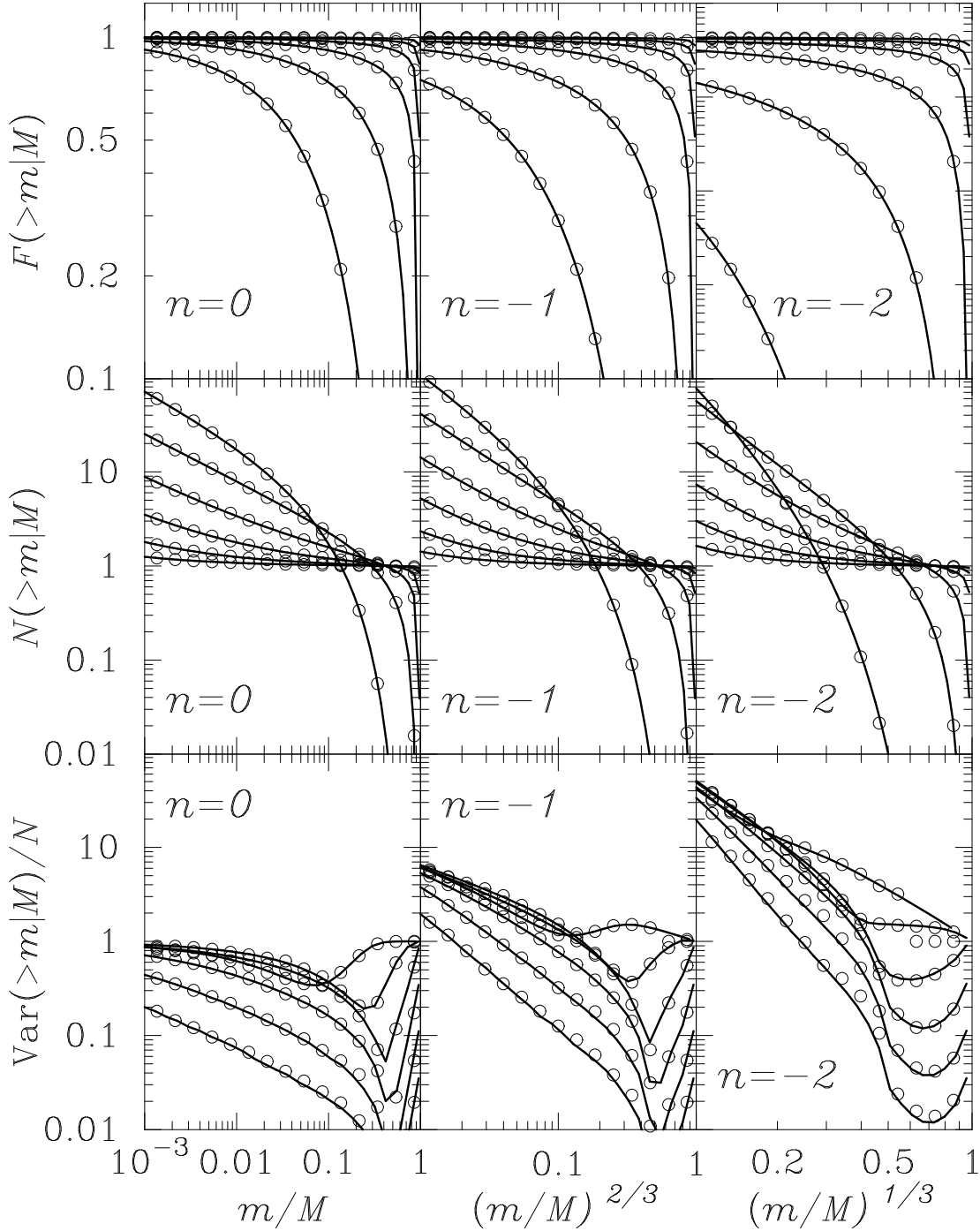}}
\caption{Results from using the first algorithm described in the 
text to partition an $M=M_*$ halo into subclumps at an earlier time, 
specified by $(z_1-z_0)= 0.01$. $0.03$, $0.1$, $0.3$, $1$ and $3$.  
In the top and middle panels, curves that are higher on the right 
correspond to smaller values of $(z_1-z_0)$.  
The trend is the opposite in the bottom panel.
Symbols show the Monte-Carlo results, curves show analytic
results (from \ref{fmMps}, \ref{nmM} and \ref{c120p}).}
\label{epsmom}
\end{figure*}

\section{Monte-Carlo realizations of merger history trees}
\label{mctree}
In this section, the two partition algorithms discussed in the 
previous section are used to produce ensembles of partitions.  
These partitions are compared with the analytic results described 
earlier.  
The second algorithm is then embedded into a loop over small 
timesteps to generate ensembles of merger history trees.  
Various properties of these trees are also compared with the 
analytic formulae.  
Comparison with halos in N-body simulations is the subject of 
the next section.  

\begin{figure*}
\centering
\mbox{\psfig{figure=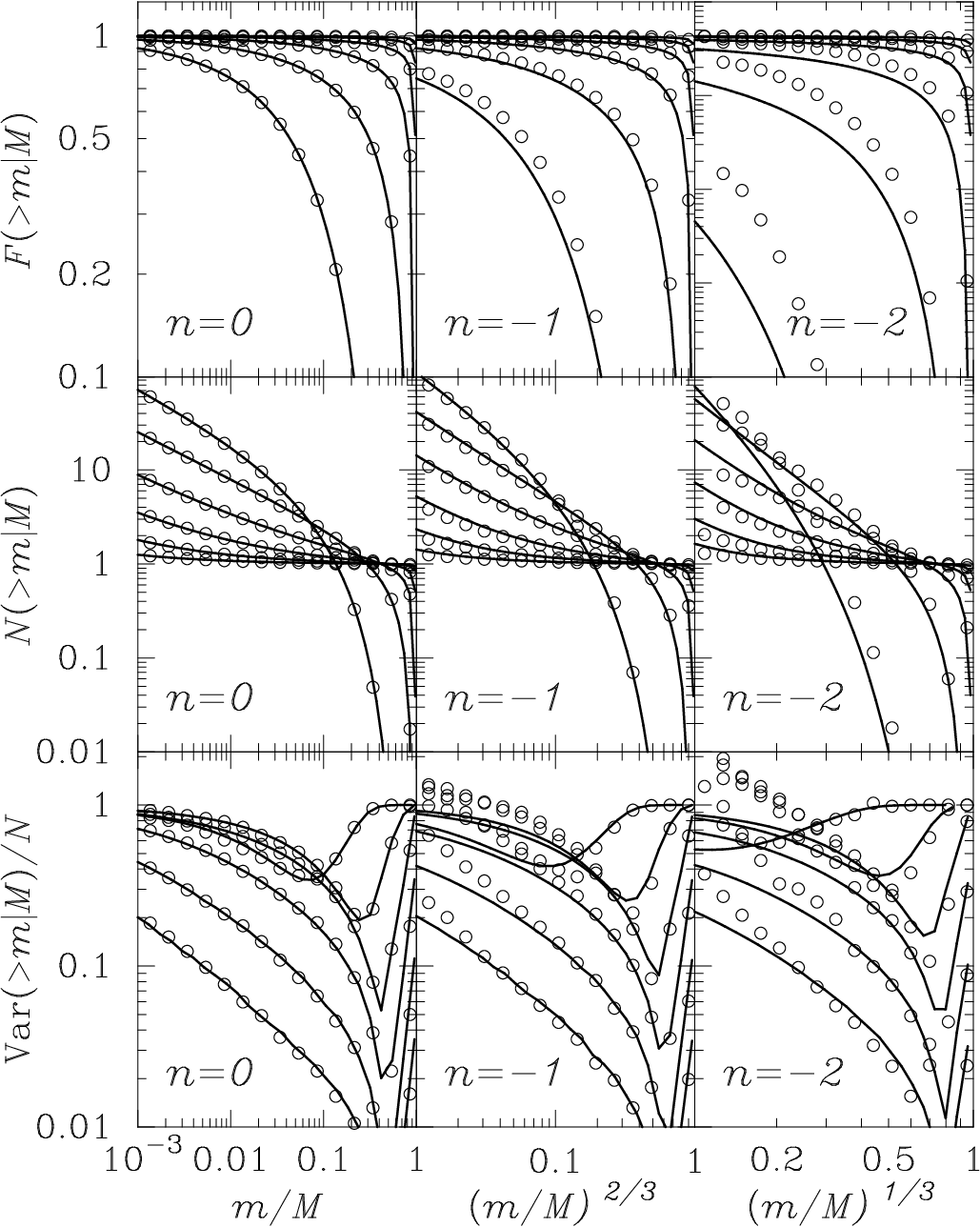,height=18.5cm}}
\caption{Results from using the second algorithm described in the 
text to partition an $M=M_*$ halo into subclumps at an earlier time, 
specified by $(z_1-z_0)= 0.01$. $0.03$, $0.1$, $0.3$, $1$ and $3$. 
In the top and middle panels, curves that are higher on the right 
correspond to smaller values of $(z_1-z_0)$.  
The trend is the opposite in the bottom panel.  
Symbols show the Monte-Carlo results, curves show analytic 
formulae (from \ref{fmMps}, \ref{nmM} and~\ref{c120}) for comparison.}
\label{epsalt}
\end{figure*}

We show results for partitions associated with initially 
scale free power-spectra:  $P(k)\propto k^n$.  In this case, 
$\sigma^2(m)\propto m^{-\alpha}$, where $\alpha=(n+3)/3$.
For such power-spectra, expressions like $N(m|M)$ and $c(mm|M)$ 
are functions of two parameters only:  
$m/M$ and $(\delta_1-\delta_0)/\sigma(M)$.  
It is convenient to define 
$\delta = \delta_{\rm c}(1+z) = \sigma(M_*)(1+z)$, so that 
$(\delta_1-\delta_0)/\sigma(M) = (z_1-z_0)\,(M/M_*)^{\alpha/2}$. 
The curves below are plotted as functions of $(m/M)^\alpha$.  
They were produced for a range of choices for $(z_1-z_0)$, 
with $M$ set equal to $M_*$.  The scaling above shows that by 
rescaling $(z_1-z_0)$ appropriately, the same curves represent 
other values of $M/M_*$.  For example, if $M=\mu M_*$, then these 
same curves represent the subclump distribution at 
$(z_1-z_0)/\mu^{\alpha/2}$.  

Figs.~\ref{epsmom} and~\ref{epsalt} show results for initial 
power spectra with slopes $n=0$, $-1$, and $-2$.  
In all panels, the thick solid curves show the analytic 
formulae with $M=M_*$ and $(z_1-z_0) = 0.01$, $0.03$, $0.1$, 
$0.3$, $1$ and $3$.  Symbols show the corresponding quantities, 
averaged over 4,000 random partitions of $M$, 
generated using the algorithms described above.  
The different panels in each figure show the cumulative 
quantities $F(>\!m|M)$, $N(>\!m|M)$, and Var$(>\!m|M)/N(>\!m|M)$.  
In both figures, $F(>\!m|M)$ and $N(>\!m|M)$ are got by integrating 
the excursion set equations~(\ref{fmMps}) and~(\ref{nmM}) from 
$m$ to $M$.  Var$(>\!m|M)/N(>\!m|M)$ is the variance in the number 
of subclumps more massive than $m$, divided by the mean, $N(>\!m|M)$; 
if the number of subclumps were Poisson distributed among the 
halos, this quantity would be unity.  
The solid curve for Var$(>\!m|M)/N(>\!m|M)$ in Fig.~\ref{epsmom} 
is got by integrating equation~(\ref{c120p}) over the 
relevant range in $m_1$ and $m_2$, doing the necessary 
integral to get the mean $N(>\!m|M)$, so computing the 
variance, and then dividing by the mean.  
The corresponding curve in Fig.~\ref{epsalt} is got by 
integrating equation~(\ref{c120}) (instead of~\ref{c120p}) over 
the necessary range in $m$.  

\begin{figure*}
\centering
\mbox{\psfig{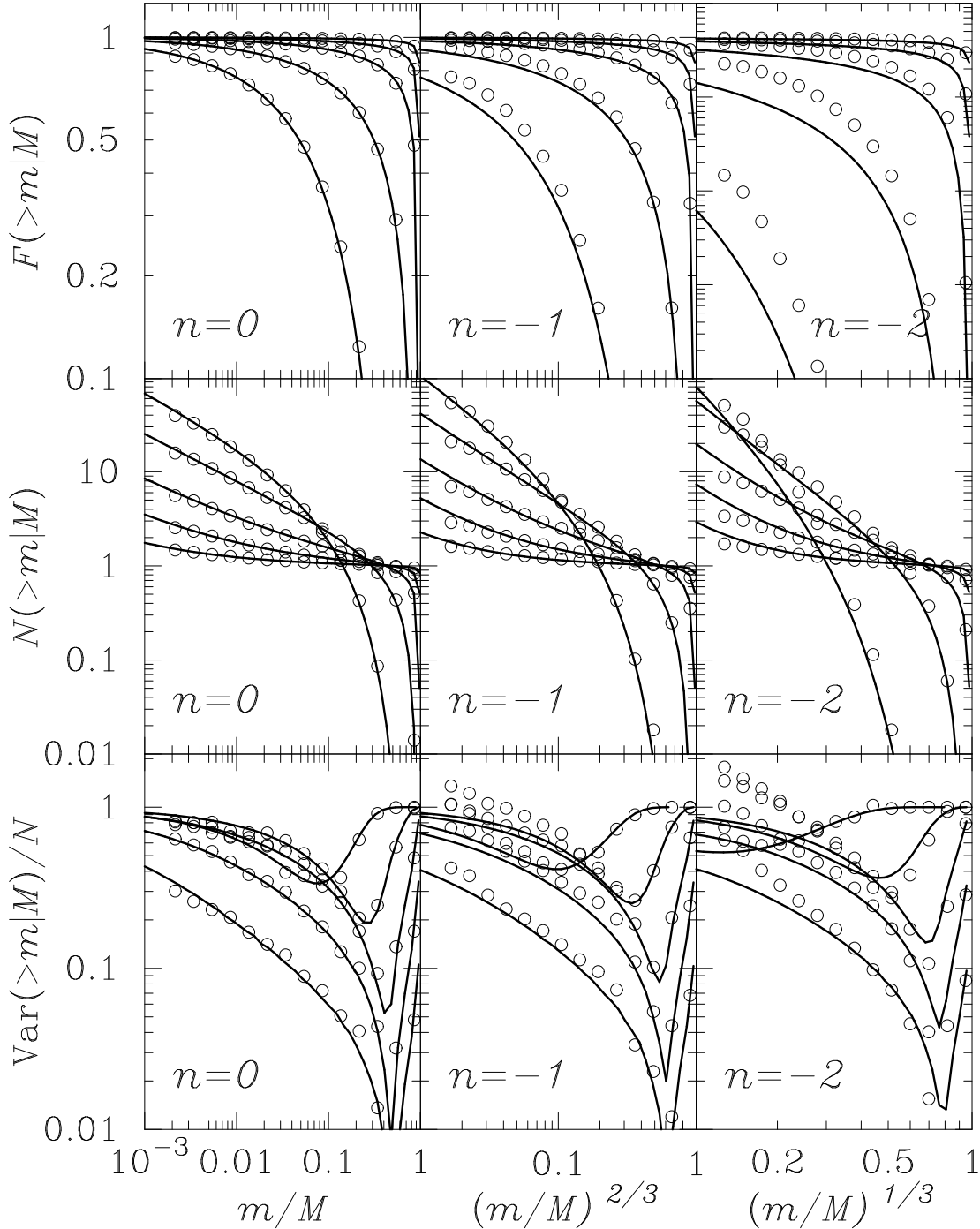}}
\caption{Results from embedding the second algorithm described in 
the text into a loop over small time-steps ($\Delta z=0.01$) to 
generate the merger history tree of an $M=M_*$ halo.  
Figure shows quantities at an earlier time, specified by 
$(z_1-z_0)= 0.03$, $0.1$, $0.3$, $1$, and $3$.  
Symbols show the Monte-Carlo results, curves were made 
similarly to those in Fig.~\ref{epsalt}.}
\label{epsrecur}
\end{figure*}

The symbols in Fig.~\ref{epsmom} are well fit by our analytic 
formulae.  This shows that the algorithm works as planned.  
The fits to the mean excursion set quantities (top two panels) 
are excellent, as is the fit to the second order moment 
(bottom panel). 

The ratio of the variance to the mean almost always has a 
minimum---this is a consequence of mass conservation.  
Consider the range $m/M>0.5$ for a given redshift interval 
$z\equiv z_1-z_0$.  
Let $p(n,z)$ denote the probability that $M$ has $n$ subclumps 
in this range.  
Mass conservation requires that there be at most one subclump in 
this range.  Therefore $p(n,z)=0$ for all $n>1$, so the mean is 
$\mu_1 = p(1,z)$, the variance is $\mu_2 = p(1,z)-p(1,z)^2$, and 
the ratio $\mu_2/\mu_1$ is $1-p(1,z)$, which is always less than 
unity.  Since $p(1,z)$ refers to all subhalos that are more massive 
than $m$, it will certainly not decrease as $m/M$ decreases from 
$1$ to $0.5$ (since decreasing $m/M$ corresponds to increasing the 
allowed range of subhalo masses).  Therefore, over this range, 
$\mu_2/\mu_1$ will decrease as $m/M$ decreases.  
The rate of decrease will depend on the redshift interval, 
since, for a given $m/M>0.5$, $p(1,z)$ will decrease as $z$ 
increases.  (This just reflects the fact that it is much less likely 
that $m/M\approx 1$ at, say, $z=5$ than at, say, $z=0.5$.)  
When $m/M\ll 1$, then the restriction implied by mass conservation 
becomes less severe, so the scatter in the subhalo counts can 
increase, and may even exceed the Poisson value.  
Fig.~\ref{epsmom} shows that, for white noise, this Poisson 
value is never exceeded.  
The ratio sometimes exceeds unity for other power-spectra, 
at small masses.  This is because, for this algorithm, 
subclumps of mass $\mu$ always come in groups of $\nu>1$, 
so it is as though less massive clumps are clustered, and 
so the variance in the counts of these clumps exceeds the 
mean.  

\begin{figure*}
\centering
\mbox{\psfig{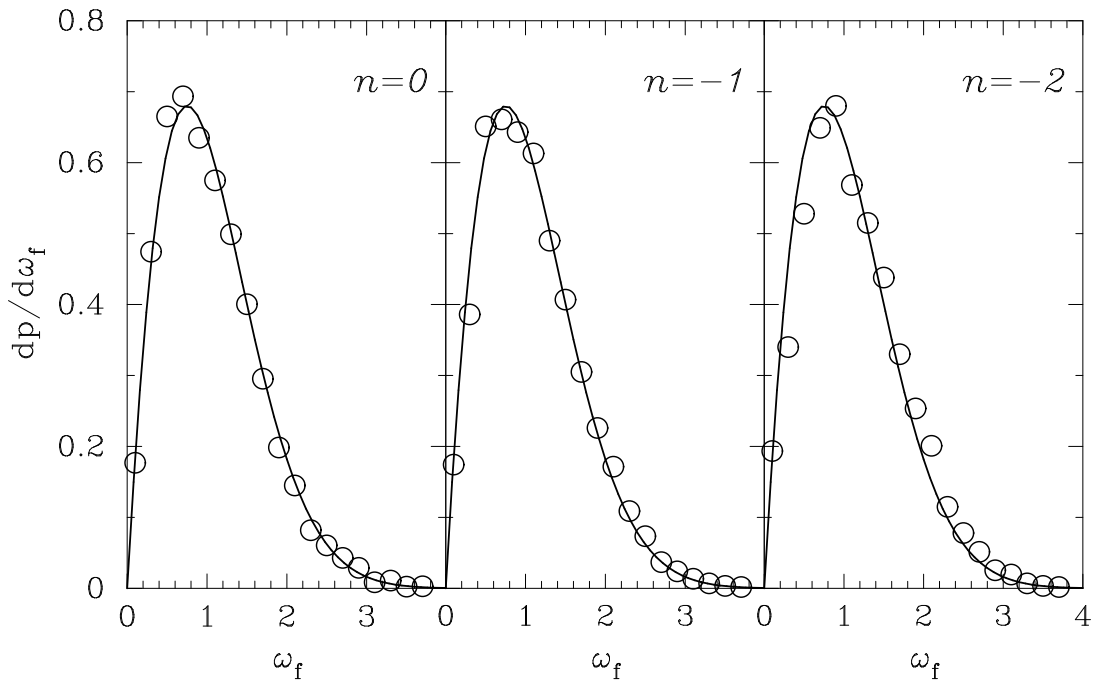}}
\caption{The scaled distribution of formation times.  Open circles 
show the distribution generated using our merger tree algorithm, 
solid lines show equation~(\ref{eq233}), originally derived by 
Lacey \& Cole (1993).  The solid lines provide good fits to the 
distribution measured in numerical simulations (Lacey \& Cole 1994).}
\label{form}
\end{figure*}

The symbols in Fig.~\ref{epsalt} show the corresponding results 
generated using our second algorithm (note that the range shown 
on the $y$-axis in the bottom panel is not the same as in the 
previous figure).  For $n=0$, of course, there is no difference 
between the two algorithms.  For the other values of $n$, the 
algorithm does not reproduce the excursion set mean 
values exactly, though for $m/M\ge 0.01$ or so the agreement 
bewteen the symbols and the curves is fair.  Over this mass 
range, the variance given by equation~(\ref{c120}) provides 
a fairly good description of the variance in the Monte-Carlo 
ensemble.  The algorithm generates partitions with slightly 
too many massive subclumps relative to less massive subclumps.  
For $n\ne 0$, this means that the discrepancy between the 
algorithm and the theoretical curves is more obvious in the 
plots of $F(>\!m|M)$ than in those of $N(>\!m|M)$.  

Fig.~\ref{epsrecur} shows the result of generating an 
ensemble of realizations of the merger history tree of an 
$M=M_*$ halo by embedding the second of our algorithms into a 
loop over small time steps ($\Delta z=0.01$).  
The figure shows the same cumulative distributions as before.  
Symbols show the Monte-Carlo results, and curves were 
made similarly to those in Fig.~\ref{epsalt}.  

For $n=0$, the symbols are well fit by the curves, and the quality 
of the fit does not depend on $\Delta z$.  This has the following 
consequence.  Recall that the first of our algorithms uses the 
white-noise partition of $M$ into subregions, and then scales 
the number of subclumps associated with each subregion by some 
factor that depends on power spectrum.  Since the white-noise 
algorithm works even when embedded in a timeloop, with 
appropriate care, our first algorithm, when embedded 
in a loop over time steps, will continue to be described by 
the same formulae that were used in Fig.~\ref{epsmom}.  
Therefore, we have not included a figure showing this explicitly.  

Fig.~\ref{epsrecur} shows that when $n\ne 0$, merger trees 
generated using the second algorithm are not well fit by the 
analytic formulae.  This is hardly surprising, since the 
one-step partitions did not fit these formulae either, 
and the tree is made by stringing together many one-step 
partitions.  Since the difference between the symbols and the 
curves in Fig.~\ref{epsalt} depended on $(z_1-z_0)$, we might 
reasonably expect that the Monte-Carlo symbols in 
Fig.~\ref{epsrecur} will depend on the choice of $\Delta z$.  
We have explored this possibility:  for the range 
$0.001\le \Delta z\le 0.05$, this dependence is negligible.  
This is reassuring, particularly because this means that, 
at least over the mass range $m/M\ge 0.01$ or so, 
equation~(\ref{c120}) provides a good analytic approximation 
to the true variance of the subclump distribution.   
 
\section{Comparison with numerical simulations}\label{scfsim}
This section compares the ensemble of trees generated by 
embedding the second of our two partition algorithms in a loop 
over time steps with that measured in numerical simulations of 
clustering from scale-free initial conditions.  First we 
argue that the distribution of halo formation times generated 
by our merger tree algorithm is in reasonable agreement with 
what was measured by Lacey \& Cole (1994) in their simulations.  
Then we consider a variety of other statistical tests, and 
compare the results generated using our merger tree algorithm 
with those which occured in scale-free simulations kindly 
made available by S. White.  

Suppose we consider the mass of the largest subclump of the 
largest subclump of the largest subclump and so on.  
Define the formation time $z_{\rm f}$ of an $M_0$ halo as the 
first time that this mass drops below half of $M_0$.  
Lacey \& Cole (1993) showed that the distribution of formation 
times defined in this way, suitably rescaled, depends only weakly 
on the power spectrum.  They showed that, for white-noise initial 
conditions, this scaled distribution is 
\begin{equation}
{{\rm d}p/{\rm d}\omega_{\rm f}} = 
2\,\omega_{\rm f}\,{\rm erfc}(\omega_{\rm f}/\sqrt{2}),
\label{eq233}
\end{equation}
where 
$\omega_{\rm f}\equiv z_{\rm f}\,(M_0/M*)^{\alpha/2}/\sqrt{2^\alpha - 1}$.  
Lacey \& Cole (1994) showed that this distribution 
provides a good fit to the scaled formation time distribution 
measured in numerical simulations of hierarchical clustering.  
The solid curves in each panel of Fig.~\ref{form} show this 
analytic formula.  
The open circles in Fig.~\ref{form} show the corresponding 
scaled distribution of formation times (averaged over $10^4$ 
realizations of the merger tree) generated by our merger tree 
algorithm.  The symbols are reasonably well described by 
the formula.  Since the formula describes the simulation 
results reasonably well also, we conclude that our merger tree 
algorithm generates formation time distributions that are similar 
to those in simulations.  

Next, we will consider a variety of other statistical tests;  
some are variants of those used by Kauffmann \& White (1993), 
and others are variants of those used by Sheth (1996) and 
Sheth \& Pitman (1997).  
The numerical simulations used here are the same as those studied 
by Mo \& White (1996), where they are described in more detail.  
They are normalized so that the number of particles in an 
$M_*$ halo is $M_*(a) = (a/\delta_{\rm c0})^{6/(n+3)}$, 
where $a$ is the expansion factor since the initial time, and 
$n$ is the slope of the initial power spectrum.  
We show results for $n=0$ and $n=-1.5$ below.  
We will sometimes quote results in terms of $M_*$:  
for all figures below, we use $\delta_{\rm c0}=1.7$, and we 
study the merger trees of $M_0$ halos identified at a time 
when an $M_*$ halo contains $471$ and $166$ particles for 
$n=0$ and $-1.5$, respectively.  

\begin{figure}
\centering
\mbox{\psfig{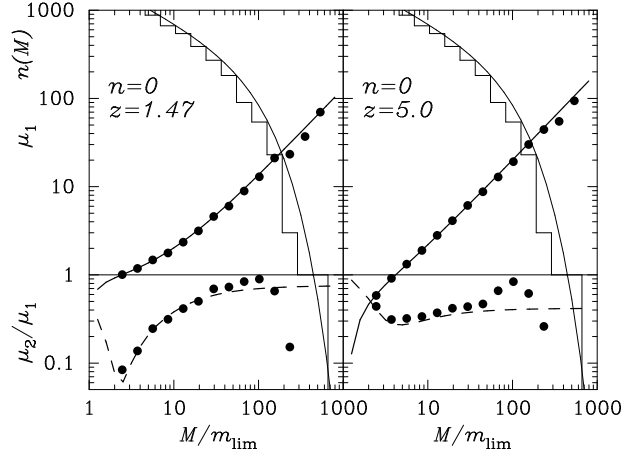}}
\caption{The mean $\mu_1$, and the variance divided by the mean 
$\mu_2/\mu_1$, of the number of subclumps that, at $z$, have mass 
greater than $m=10$ that are within a halo with mass $M/m$.  
Curves show the analytic results; for $n=0$ these curves 
describe the results of our merger tree algorithm exactly.  
The filled circles show the corresponding 
quantities measured in the $n=0$ simulations.  
The histogram shows the unconditional mass function measured in 
the simulations, and the thin solid curve shows the associated 
Press--Schechter mass function.  }
\label{pnmom0}
\end{figure}

\begin{figure}
\centering
\mbox{\psfig{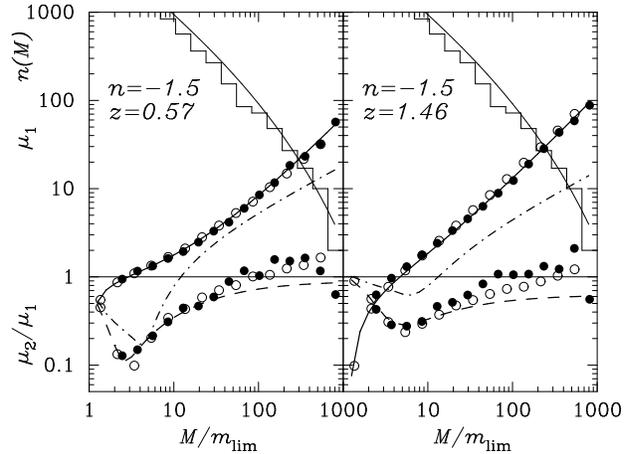}}
\caption{The same as the previous figure, but for $m=5$ and $n=-1.5$.  
Filled circles show the simulation results, 
open circles show the corresponding quantities generated using 
our tree algorithm, curves show the analytic formulae, since these 
formulae do not describe the tree results exactly when $n\ne 0$.}
\label{pnmom1h}
\end{figure}

In what follows we choose a minimum mass threshold $m$.  
At a given output time $z_0$ we identify all halos with mass 
$M\ge m$.  In the remainder of this section, $z_0\equiv 0$.  
At an earlier output time $z_1>z_0$ we identify 
all subclumps of each halo identified at $z_0$ that, at $z_1$ 
are more massive than $m$.  For each halo, we store the number of 
such subclumps as well as its square.  Averaging over all $z_0$ 
halos that have the same $M/m$ allows us to compute the mean and 
the variance of the cumulative subclump distribution.  
In the previous section we showed these quantities as functions 
of $m/M$ as $m$ varied at fixed $M$, for representative values 
of $M$ and $z_1-z_0$.  In Figs.~\ref{pnmom0} and~\ref{pnmom1h}, 
we plot these quantities versus $M/m$ for fixed $m$ as $M$ varies.  
In all the plots below, when $z_0=0$, then $m/M_*=10/471$ for 
the $n=0$ simulations, and $m/M_*=5/167$ for $n=-1.5$.  

Figs.~\ref{pnmom0} and~\ref{pnmom1h} show the first two moments 
of the subclump distribution as a function of $M/m$.  
The histogram shows the number of halos in each mass bin 
that were found in the simulations.  
The thin solid curves in show the expected number 
computed using the unconditional mass function 
$n(M)\,{\rm d}M\propto f(M)\,{\rm d}M/M$ of 
equation~(\ref{fmps}).  There are two reasons for showing 
this histogram.  This first is to demonstrate that the unconditional 
mass function is in reasonable agreement with the simulations.  
The second is that the histogram is intended 
to give an idea of the mass range over which the 
halo sample in the simulations is approximately 
statistically complete (e.g., some of the statistical quantities 
measured in the simulations may not be accurate measures of the 
true quantities in the mass range where the number of halos in 
a given mass bin is small.)

The thick solid curves show the conditional mass function 
$\mu_1\equiv N(>\!m|M)$ computed by doing the necessary integral 
over equation~(\ref{nmM}).  The thick dashed curves show 
$\mu_2/\mu_1\equiv{\rm Var}(>\!m|M)/N(>\!m|M)$ computed using 
equation~(\ref{c120}), and the thick dot--dashed curves show 
this same quantity computed using equation~(\ref{c120p}).  
(For $n=0$ these two curves overlap exactly.)  
Recall that the curves in Figs.~\ref{epsmom} and~\ref{epsalt} 
depend on the combination $(z_1-z_0)^2 S_*/S$.  Although the 
curves there were made for $M=M_*$ halos, if $M/M_*$ is different, 
then the same curves represent a different value of $z_1-z_0$ 
for scale-free initial conditions.  So, with a little thought, 
one can convince oneself that the curves there are consistent 
with those shown here.  


The filled circles show the corresponding quantities measured in 
the numerical simulations.  Open circles show these quantities 
generated by our merger history algorithm.  When $n=0$, the tree 
results are fit by the analytic formulae exactly, so we only show 
the theoretical curves.  When $n=0$, these curves fit the simulation 
results reasonably well.  
When $n=-1.5$, the merger tree algorithm is not fit exactly by 
the analytic curves, and the two analytic curves for the variance 
differ from each other.  Figure~\ref{pnmom1h} shows that the 
numerical simulation results are fit better by 
equation~(\ref{c120}) than by~(\ref{c120p}), and still better by 
the results generated by our algorithm.  

\begin{figure}
\centering
\mbox{\psfig{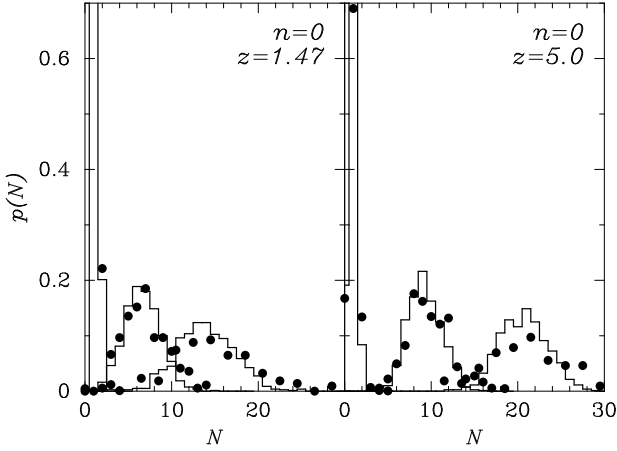}}
\caption{The distribution of the number of subclumps that, at $z$, 
have mass greater than $m/M_*=10/471$ that are within a halo with 
mass $M_0/M_*=38/471=0.08$ (left-most), $447/471=0.95$ (middle) and 
$1024/471=2.17$ (right-most).  
Histograms show the distribution measured using our algorithm, 
and filled circles show the corresponding distribution measured 
in the $n=0$ simulations.  }
\label{pnp0}
\end{figure}
\begin{figure}
\centering
\mbox{\psfig{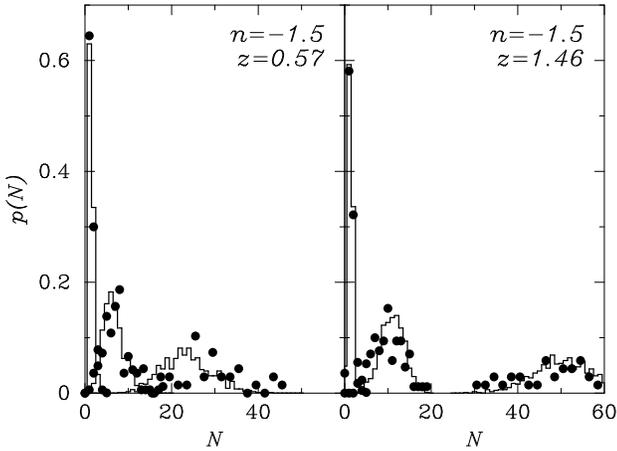}}
\caption{The distribution of the number of subclumps that, at $z$, 
have mass greater than $m/M_*=5/166$ that are within a halo with 
mass $M_0/M_*=28/166=0.17$ (left-most), $339/166=2.04$ (middle) and 
$1780/166=10.7$ (right-most).  
Histograms show the distribution measured using our algorithm, 
and filled circles show the corresponding distribution measured 
in the $n=-1.5$ simulations.}
\label{pnp1h}
\end{figure}
\begin{figure}
\centering
\mbox{\psfig{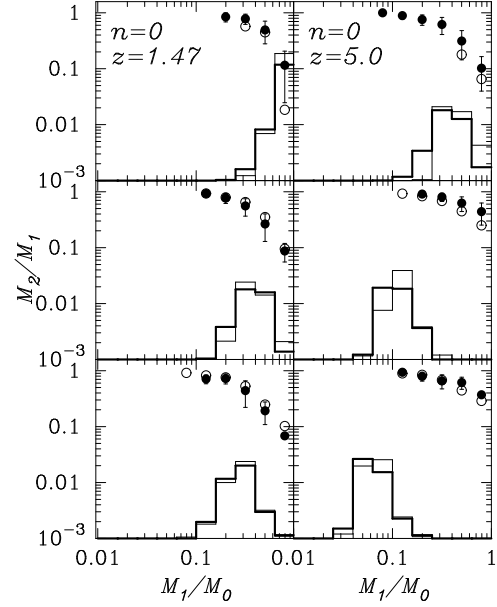}}
\caption{The relative sizes of the first and second largest 
subclumps, $M_1/M_0$ and $M_2/M_1$, for the same three values 
of $M_0$ as in Fig.~\ref{pnp0}.  
Thin histograms show the distribution of $M_1/M_0$ measured using 
our algorithm, and bold histograms show the corresponding 
distribution measured in the $n=0$ simulations.  Open circles 
in the top right show the mean of $M_2/M_1$ given $M_1/M_0$ 
generated by our algorithm; filled circles show the corresponding 
mean and rms scatter measured in the simulations.  }
\label{eml0}
\end{figure}
\begin{figure}
\centering
\mbox{\psfig{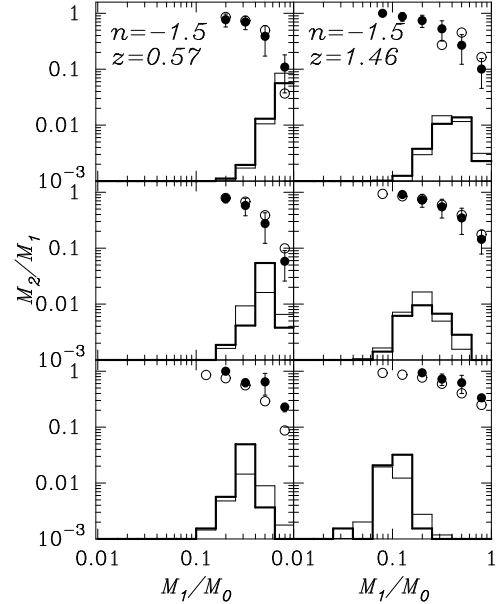}}
\caption{Same as the previous figure, but for $n=-1.5$, and for 
the same three choices of $M_0$ shown in Fig.~\ref{pnp1h}.  }
\label{eml1h}
\end{figure}

Figs.~\ref{pnp0} and~\ref{pnp1h} show the distribution of the number 
of progenitors of an $M_0$ halo that, at $z$, are more massive than 
$m$.  The histogram shows this quantity generated by averaging over 
800 realizations of the merger history tree using our algorithm, 
and solid points show the corresponding quantity measured in the 
simulations.  For both values of $n$ studied, three representative 
values of $M_0$ are shown:  these approximately bracket the range 
$0.1\le M/M_*\le 10$.  For both values of $n$, the histograms 
generated by our algorithm are in reasonable agreement with the 
numerical simulation results.  

Figs.~\ref{eml0} and~\ref{eml1h} show the relative sizes of the 
subclumps produced by our algorithm and those measured in the 
simulations.  The histograms show the distribution of the size 
of the largest subclump $M_1/M_0$ for the same values of $M_0$ 
as in the previous figures.  (The topmost panels show the lowest 
values of $M_0$.)  The symbols show the corresponding simulation 
results.  The open circles in the top right show the mean of 
$M_2/M_1$ given $M_1/M_0$, where $M_2$ denotes the size of the 
second largest subclump, averaged over 800 realizations of the 
merger tree.  The solid circles and error bars show the associated 
quantities measured in the simulations.  For both values of $n$ 
our algorithm is in good agreement with the simulations, though 
the agreement when $n=0$ is somewhat better than when $n=-1.5$.  

We conclude this section with the observation that, for most of 
the tests we considered, our algorithm produces forests of 
merger history trees that are reasonably similar to those which 
occur in numerical simulations of hierarchical clustering.  

\section{Discussion}\label{concl}
We described an algorithm which allowed us to partition 
halos at a given time into subhalos at an arbitrary 
earlier time.  The algorithm is exact for Poisson and 
white-noise initial conditions.  In these cases, we 
provided analytic expressions for the higher order moments 
of the subclump distribution.  We discussed two possible 
ways of modifying the algorithm to partition halos which form  
from more general Gaussian initial conditions.  
We then showed that the one-step partition algorithm can 
be embedded into a loop over many time steps to generate 
a forest of merger history trees.  This also showed why binary 
split algorithms previously used to construct the merger trees 
may be reasonably accurate.  

We then embedded the second of our partition algorithms into a 
loop over time steps to generate the forest of merger trees.  
We compared this forest of trees with the ensemble of trees 
measured in numerical simulations of hierarchical gravitational 
clustering from scale-free initial conditions.  
For the initial conditions we studied, the agreement between our 
trees and those of the simulations was fair.  Furthermore, this 
comparison showed that our analytic formulae for the higher order 
moments of the subclump distribution provided reasonable fits to 
the numerical simulation results even when the initial conditions 
were different from white noise.  

This result is particularly useful for studying the spatial 
distribution of dark matter halos.  
Mo \& White (1996) argued that the higher order moments of the 
subclump distribution associated with the forest of merger history 
trees of dark matter halos can be related to the higher order moments 
of the spatial distribution of these halos.  
Since the higher order moments associated with our merger 
tree algorithm, and also of the merger trees in the numerical 
simulations, were reasonably well fit by our equation~(\ref{c120}), 
it is possible to write down analytic approximations to spatial 
quantities like the halo--halo correlation function that are 
also reasonably accurate.  
This is the subject of ongoing work (Sheth \& Lemson 1998).  

Before concluding, we note that Sheth (1998) describes a model for 
clustering from compound Poisson initial conditions.  
For these more general initial conditions, the analogue of 
equation~(\ref{fjk}) factors similarly to how it does for the 
Poisson case.  This means that a similar algorithm for generating 
the associated forest of merger trees can be used there also.  

\section*{Acknowledgments}
Thanks to Jim Pitman for discussing his results on 
size--biased distributions which, in the white-noise case, 
lead to this same partition algorithm.

\appendix

\section{The partition probability function}\label{ppf}
This Appendix describes a simple interpretation of the 
partition probability function (equation~\ref{prtfnc}).  
It shows that this partition formula is consistent with the 
assumption that mutually disconnected volumes in a Poisson 
distribution are mutually independent.  

Since $b_0$ can be related to a density (equation~\ref{bdt}), 
an $(M,b_0)$-halo is associated with a region in the initial 
Poisson distribution that has size 
\begin{equation}
_0\!V_M =  M/[\bar n\,(1+\delta_0)] = Mb_0/\bar n,
\end{equation}
where $\bar n$ denotes the average density.  
This means that $p(n_1\cdots n_M,b_1|M,b_0)$ of 
equation~(\ref{prtfnc}) denotes the probability 
that a region containing exactly $M$ particles at average density 
$\bar n(1+\delta_0)$ has $n$ subregions, each with average 
density $\bar n(1+\delta_1)$, where $\delta_1\ge\delta_0$, 
of which $n_1$ such subregions contain only one particle 
(so they each have size $_1\!V_1$), $n_2$ contain two particles 
(so they each have size $_1\!V_2$), etc.  
Notice that the sum of the sizes of these $n$ sub-volumes 
is $_1\!V_M <\, _0\!V_M$, since $b_1\le b_0$.  The remaining 
volume $_0\!V_M - _1\!V_M = M(b_0-b_1)/\bar n$ is empty.  

Now suppose that the region $_0\!V_M$ containing $M$ 
particles is known to contain $n$ $b_1$-subregions, of 
which at least one such subregion has size $_1\!V_m$ and contains 
exactly $m$ particles.  The probability that $_0\!V_M$ contains 
$n_1$ $b_1$-regions of size $_1\!V_1$, $n_2$ $b_1$-regions of 
size $_1\!V_2$, etc., given that, within $_0\!V_M$, there 
is at least one $b_1$-subregion of size $_1\!V_m$ 
containing $m$ particles, and $n-1$ other $b_1$-subregions is 
\begin{equation}
{(Mb_{01})^{n-1}\,{\rm e}^{-Mb_{01}}\over \eta(M,b_0)}\,
{\eta(m,b_1)\over \langle n_m,b_1|M,b_0\rangle}
\prod_{i=1}^{M-m} {\eta(i,b_1)^{n_i}\over n_i!},
\label{eq2}
\end{equation} 
where $b_{01} = b_0-b_1$, $\sum_i n_i = n-1$, 
and $\sum_i i\,n_i = M-m$.  The term 
\begin{equation}
\langle n_m,b_1|M,b_0\rangle = 
\sum_{\rm all\ parts.} n_m\, p(n_1\cdots n_M,b_1|M,b_0) ,
\end{equation}
which is the sum over all partitions of $M$ that have at 
least one $m$-part, is included in the denominator since this 
is a conditional probability.  However, equation~(\ref{facij}) 
shows that 
\begin{eqnarray}
\langle n_m,b_1|M,b_0\rangle = 
Mb_{01}\,{\eta(m,b_1)\,\eta(M-m,b')\over \eta(M,b_0)},
\label{eq3}
\end{eqnarray} 
where $b'$ is defined by the relation 
\begin{equation}
{M-m\over _0\!V_M -\, _1\!V_m} \equiv \bar n(1+\delta') 
= {\bar n\over b'} .
\label{bprm}
\end{equation}  
Thus, $b'$ parameterizes the density in the remaining volume 
$_0\!V_M -\, _1\!V_m$.  Substituting equation~(\ref{eq3}) for 
$\langle n_m,b_1|M,b_0\rangle$ into~(\ref{eq2}) yields 
\begin{equation}
{(Mb_{01})^{n-2}\,{\rm e}^{-Mb_{01}}\over \eta(M-m,b')}\,
\prod_{i=1}^{M-m} {\eta(i,b_1)^{n_i}\over n_i!} .
\label{eq6}
\end{equation}
However, 
\begin{equation}
(M-m)\,(b'-b_1) = M(b_0-b_1) = Mb_{01},
\label{kb21}
\end{equation}
so~(\ref{eq6}) can be written as 
\begin{equation}
{[(M-m)(b'-b_1)]^{n-2}{\rm e}^{-(M-m)(b'-b_1)}\over \eta(M-m,b')}\,
\prod_{i=1}^{M-m} {\eta(i,b_1)^{n_i}\over n_i!},
\end{equation}
where $\sum_i n_i = n-1$, and $\sum_i i\,n_i = M-m$.  
Clearly, this is the same as 
\begin{equation}
p(n_1\cdots n_{M-m},b_1|M-m,b') ,
\end{equation}
where the $n_i$s sum to $(n-1)$, and the sum of $in_i$ is $(M-m)$.  

This shows explicitly that if the region $_0\!V_M$ containing 
exactly $M$ particles, so the average density within it is 
$M/ _0\!V_M=\bar n(1+\delta_0)$, is known to contain exactly 
$n$ subregions, each with average density $\bar n(1+\delta_1)$, 
and it is also known that one of these subregions has size 
$_1\!V_m$ and contains exactly $m$ particles, 
so the remaining $M-m$ particles are arranged into $n-1$ 
subregions, each with the same overdensity $\delta_1$, in 
the remaining volume $_0\!V_M -\, _1\!V_m$, then the 
distribution of sizes of the $n-1$ subregions (each with 
overdensity parameter $b_1$) within $_0\!V_M -\, _1\!V_m$ is 
given by the same partition probability function as for the total 
volume $_0\!V_M$, with trivial adjustments for the fact that the 
average density within the remaining volume $_0\!V_M -\, _1\!V_m$ 
will be different from that within $_0\!V_M$ itself, and for the 
fact that we need now only consider partitions of $M-m$ into 
$n-1$ parts, rather than of $M$ into $n$ parts.  
In other words, if $p({\bmath n},b_1|M,b_0)$ is viewed as 
describing the distribution of $b_1$-subregions within 
the region $_0\!V_M$, then it is fully consistent 
with the fact that non-overlapping volumes in a Poisson 
distribution are mutually independent.  

This property of the partition formula could have been inferred 
directly from the additive coalescent interpretation of 
equation~(\ref{prtfnc}) that is provided by Sheth \& Pitman (1997).  

\section{Summing over permutations}\label{permute}
Consider a partition ${\bmath m}$ of $M$ into $n$ parts:  
${\bmath m}\equiv (m_1,\cdots,m_n)$, where $\sum_i^n m_i = M$.  
Let $R_i\equiv M-\sum_{j=1}^i m_j$.  
This Appendix shows that the sum over all permutations 
$\pi[{\bmath m}]$ of $(m_1,\cdots,m_n)$, 
of $\prod_{i=1}^n 9m_i/R_{i-1})$ is unity.  This result 
is used in the main text to simplify equation~(\ref{perms}).  

Before proceeding, it is useful to define 
\begin{equation}
r_{i\cdots j\cdots k}\equiv M-m_i-\ldots -m_j-\ldots -m_k.  
\end{equation}
To illustrate the argument, suppose that $n=3$.  
Since $M = m_1+m_2+m_3$, 
\begin{equation}
{m_i\over r_{jk}} = 1 \qquad{\rm and}\qquad 
{(m_i+m_j)\over r_k} = 1 ,
\label{mijk}
\end{equation}
for all combinations of $1\le i,j,k\le 3$ in which $i\ne j\ne k$.  
Explicitly, the sum over the $n! = 6$ permutations of 
$(m_1,m_2,m_3)$ is 
\begin{displaymath}
{m_1\over M}{m_2\over r_1}{m_3\over r_{12}} + 
{m_1\over M}{m_3\over r_1}{m_2\over r_{13}} 
+ {m_2\over M}{m_1\over r_2}{m_3\over r_{21}} 
+ {m_2\over M}{m_3\over r_2}{m_1\over r_{23}} 
\end{displaymath}
\begin{displaymath}
\qquad\qquad\qquad\qquad
\ \ \ + {m_3\over M}{m_1\over r_3}{m_2\over r_{31}} 
+ {m_3\over M}{m_2\over r_3}{m_1\over r_{32}}.
\end{displaymath}
The first identity of~(\ref{mijk}) shows that this is the same as 
\begin{displaymath}
{m_1\over M}{m_2\over r_1} + {m_1\over M}{m_3\over r_1} 
+ {m_2\over M}{m_1\over r_2} + {m_2\over M}{m_3\over r_2} 
+ {m_3\over M}{m_1\over r_3} + {m_3\over M}{m_2\over r_3} .
\end{displaymath}
But this is just 
\begin{displaymath}
{m_1\over M}{(m_2+m_3)\over r_1} + 
{m_2\over M}{(m_1+m_3)\over r_2} + 
{m_3\over M}{(m_1+m_2)\over r_3} ,
\end{displaymath}
and use of the second identity of~(\ref{mijk}) reduces this to 
\begin{displaymath}
{m_1\over M} + {m_2\over M} + {m_3\over M},
\end{displaymath}
which is unity, by definition.  

It is easy to see that, for all values of $n$, the requirement 
that $M = \sum_i^n m_i$ provides relations analogous 
to those in~(\ref{mijk}), and that repeated use of these relations 
will reduce the sum of $\prod_{i=1}^n (m_i/R_{i-1})$ over all 
$n!$ permutations of the $m_i$ to unity.  

\end{document}